\documentstyle[aps,prl,multicol,epsfig]{revtex}
\catcode`\@=11

\renewcommand{\widetext}{\end{multicols} \global\columnwidth42.5pc}

\newcommand{\be}{\begin{equation}}
\newcommand{\ee}{\end{equation}}
\newcommand{\bea}{\begin{eqnarray}}
\newcommand{\eea}{\end{eqnarray}}

\begin{document}

\baselineskip=17pt

\gdef\journal#1, #2, #3, 1#4#5#6{{#1~}{\bf #2}, #3 (1#4#5#6)}
\gdef\ibid#1, #2, 1#3#4#5{{\bf #1} (1#3#4#5) #2}

\title{Conductance and Shot Noise for Particles with Exclusion Statistics }
\author{Serguei B. Isakov$^{\dagger}$, Thierry  Martin$^{\ddagger}$, and
St\'ephane Ouvry$^{\dagger}$ }
\date{November 27, 1998}
\address{$^{\dagger}$Laboratoire de Physique Th\'eorique et Mod\`eles 
Statistiques,
B\^at. 100, Universit\'e Paris-Sud, 91405 Orsay Cedex, France}
\address{$^{\ddagger}$Centre de Physique Th\'eorique, 
Universit\'e de M\'edit\'erran\'ee
P.O. Box 907, 13288 Marseille, France}
\maketitle
\tighten{\begin{abstract}
The first quantized Landauer approach to  conductance and 
noise is generalized to
particles obeying
exclusion statistics. We derive an explicit formula for the
crossover  between the shot  and thermal noise limits and argue
that such a crossover can be used to determine experimentally
whether charge carriers in FQHE devices obey exclusion statistics. 
%\vspace{0.3cm}

%PACS numbers: 71.10.Pm, 72.70.+m, 73.40.Hm
\end{abstract}}
\begin{multicols}{2}
\narrowtext

Recent experiments measuring quantum fluctuations in the tunneling current
between two edges of a fractional Hall quantum liquid detected 
a fractional charge for the charge carriers \cite{Glattli,Nature}.
At weak tunneling, quantum shot noise 
arises from the independent tunneling events of  
quasiparticles (Schottky limit): a noise measurement constitutes 
a direct measurement of the charge \cite{Glattli}. 
However, the implications of the statistics of the 
quasiparticles on the transport and noise have remained unclear. 
Laughlin's quasiparticles  
are supposed not only to carry fractional charge, but also  to obey
fractional statistics \cite{Nobel}.
It was later argued that these quasiparticles might 
also  be described \cite{Haldane} in terms of exclusion 
statistics \cite{Haldane,ES}. 
A description of edge excitations of fractional 
quantum Hall liquids in terms of Luttinger liquids 
allowed to make  predictions on the transport and noise 
in integrable  models where the tunneling is described  by a scattering 
with point like impurities\cite{integrable}. 
But it is not clear how to extract  effects of statistics
within those models.

Our purpose here is to study the implications  of
exclusion statistics of  carriers on the quantum shot noise.
To this aim,  we use the Landauer approach \cite{Landauer} where effects 
of statistics are manifest.
The reduction of the  shot noise for fermions
due to the Pauli exclusion principle predicted within this approach 
\cite{Lesovik,Martin} was observed experimentally \cite{Pauli reduction}. 
Also, differences between the fermion and boson 
noise have been pointed out \cite{Buttiker}.
In the fractional quantum Hall effect (FQHE) experiments 
\cite{Glattli,Nature}, a reduction of noise similar to that of 
fermions was observed. 

In this Letter, we  compute   current and noise 
for exclusion statistics charge carriers in the  first quantized version 
of the Landauer approach,  the ``wave packet'' approach \cite{Martin}.
The resulting expressions are in agreement with the 
fluctuation dissipation theorem and exhibit quantum shot noise reduction.
We propose that further experiments should  
look in more detail at the crossover regime to  determine the statistics
of the charge carriers.

{\it Bose and Fermi current and noise.---}
Let us first consider for usual Bose or Fermi carriers the current and noise 
in a first quantized
Landauer approach \cite{Martin}. This is
mandatory since in the case of exclusion statistics a second quantized
formalism is not available at present.

In the Landauer setting, a current is transported
between a reservoir 1 on the left with chemical potential $\mu_1$ and a
reservoir 2 on the right with chemical potential $\mu_2$   separated by a
scattering  region of arbitrary transmission $T$ (we assume $T$ not to 
depend on the energy of carriers). The probability to have $N$ bosons in a
reservoir of chemical potential $\mu$ and in a quantum state of energy 
$\epsilon$ is $(1-x)x^N $, 
where $x\equiv e^{\beta (\mu-\epsilon)}$ is the Gibbs factor.
The probability to have  $N$ fermions is $(1+x)^{-1}x^N$ 
keeping in mind that, due to the Pauli principle, $N\ge 2$ is forbidden.
Consider a scattering event where
$N_1$ and $N_2$ particles, in a given quantum state of energy $\epsilon$,
are incident from  reservoirs 1 and 2, respectively,
and $k_1$ ($k_2$) of these  incident
particles are transmitted to reservoir 2 (1). The probability
for such an event rewrites as
(in the sequel the indices 1 and 2 always
refer  to reservoirs 1 and 2)
\be\label{prob} (1\mp x_1)^{\pm 1}x_1^{N_1} (1\mp x_2)^{\pm 1}x_2^{N_2}
{\cal{W}}^{N_1N_2}_{k_1k_2} \; ,
\ee
where the upper and lower signs stand for bosons and fermions, 
respectively, and 
\be
{\cal{W}}^{N_1N_2}_{k_1k_2}\equiv {N_1\choose k_1}{N_2\choose
k_2}T^{k_1+k_2}(1-T)^{N_1+N_2-k_1-k_2}\; .
\ee
Since $\sum_{{0\le k_1\le N_1}\atop{0\le k_2\le  N_2}}
 {\cal{W}}^{N_1N_2}_{k_1k_2} =1 $, 
the probabilities (\ref{prob}) trivially sum to 1 when all possible scattering
processes are taken into account, i.e.
$N_1,  N_2 = 0,1,2 \ldots $ in the Bose case
and $N_1,  N_2 = 0,1$ in the Fermi case.

Nothing has yet  been said  about the
probability to have  after scattering
$N_1-k_1+k_2$ particles in reservoir 1 and $N_2-k_2+k_1$ particles in
reservoir 2. Still, scattering events with
$N_1-k_1+k_2\ge 2$  or $N_2-k_2+k_1\ge 2$ should be strengthened
for bosons because of their tendency to gather in the same 
state, whereas they should be
forbidden for fermions because of the Pauli exclusion principle.

In the following, we will denote the Bose and Fermi summations as
$\sum_{{N_1=0}\atop{N_2=0}}^{\infty}
\sum_{{0\le k_1\le N_1}\atop{0\le k_2\le N_2}}\equiv \sum^{\infty}$
and $\sum_{{N_1=0}\atop{N_2=0}}^{1}
\sum_{{0\le k_1\le N_1}\atop{0\le k_2\le N_2}}\equiv \sum^1$.
The one dimensional line being oriented
 from the left to the right, the average algebraic number of carriers
 $\langle k \rangle \equiv \langle k_1-k_2 \rangle $
 of charge  $q$
(which may be different from the electron charge) at  energy  $\epsilon$
transmitted through the scattering region
---related to the average 1d current by 
$\langle I\rangle 
=(q/h)\int_{0}^{\infty}  {\langle k \rangle }\, d \epsilon$--- is
\[  
\langle k \rangle =
 \sum^{\infty , 1}
 (1\mp x_1)^{\pm 1}x_1^{N_1} (1\mp x_2)^{\pm 1}x_2^{N_2}
{\cal{W}}^{N_1N_2}_{k_1k_2}(k_1-k_2)
\]
leading  to the Landauer formula $ \langle k \rangle  =T(f_1-f_2) $
 where
 $ f(x)=x/(1 \mp x)$ is the mean ocupation
 number at energy $\epsilon$
 in the  Bose and Fermi cases.
Can the spectral noise be obtained in  this way?
The average square of the number of transmitted particles  is  
\[
\langle {k}^2 \rangle  =
\sum^{\infty , 1}
(1\mp x_1)^{\pm1}x_1^{N_1} (1 \mp x_2)^{\pm1} x_2^{N_2}
 {\cal{W}}^{N_1N_2}_{k_1k_2} (k_1-k_2)^2
\]
 to yield  the fluctuations  ---related to the zero frequency  spectral
 density of noise by
 $P =\frac{2q^2}{h}\int_0^{\infty} (\langle k^2  \rangle - 
{\langle k \rangle }^2)d\epsilon \, $---
\be\label{7} \langle k^2  \rangle - {\langle k \rangle }^2
=T(f_1+f_2)\pm T^2(f_1^2+f_2^2)  \;. \ee
However, one has to add to (\ref{7}) the term
$ \pm 2T(1-T)f_1f_2$
to recover the standard result obtained from second 
quantization\cite{Buttiker}
\bea\nonumber 
\langle k^2  \rangle - {\langle k \rangle }^2
=&T&f_1(1\pm f_1)+Tf_2(1\pm f_2)\\
\label{9} &\mp&
T(1-T)(f_1-f_2)^2  \;.  \eea
It is straightforward
to understand, in the Fermi case, the origin of  $2T(1-T)f_1f_2$
which has to be substracted
from  (\ref{7}) to get (\ref{9}): this term is precisely built by the
two scattering processes
$N_1=N_2=k_2=1$, $ k_1=0$ and $N_1=k_1=N_2=1$, $ k_2=0$,  with respectively
$N_1-k_1+k_2=2$ particles in reservoir 1 and $N_2-k_2+k_1=2$ 
particles in
reservoir 2 after scattering.  Precisely, both these processes should
be suppressed because of the Pauli exclusion principle. Why the
same term has to be added in the Bose case to take into account the
strenghtening of scattering processes where several particles occupy after
scattering  the
same energy state in reservoir 1 or 2 has yet to be understood in this first
quantized approach.

It is well known that the second quantized spectral noise defined from
(\ref{9})
satisfies two independent
constraints:

(i) Fluctuation-dissipation theorem: when $\mu_1=\mu_2$, the noise is 
proportional to the conductance, which in particular implies that the 
$T^2$ term has
to vanish when the chemical potentials are equal (thermal noise).

(ii) Quantum suppression of the noise at zero temperature: 
the noise scales as $T(1-T)$ at $\beta=\infty$ (shot noise).

{\it Exclusion statistics.---}
Let us generalize the first quantized approach to 1d 
exclusion statistics\cite{ES}
particles with statistics parameter $g$. For exclusion statistics,
the one quantum state
grand-partition function $\Xi$ at energy $\epsilon$ and chemical 
potential $\mu$ satisfies the transcendental equation
$ \Xi-x\Xi^{1-g}=1$.
The mean occupation number $f(x)\equiv
x {\Xi'\over \Xi}$ (the prime  means derivative with respect to $x$)
satisfies 
\be\label{10} x={f\over (1-g f)^{g}[1+(1-g)f]^{1-g}} \;.\ee
It interpolates between
the usual Bose and Fermi distributions when $g$ varies from $0$ to $1$.

>From $ \Xi-x\Xi^{1-g}=1$, one has  $\Xi=\sum_{N=0}^{\infty}x^NP_N$
where $P_N\equiv \prod_{k=2}^N{k-gN\over k}$.
The probability to have $N$ exclusion statistics particles in
a reservoir of chemical potential $\mu$ and in a quantum 
state of energy $\epsilon$ should thus be 
\be\label{11} {1\over\sum_{N=0}^{\infty}x^NP_N}{x^NP_N} \;.\ee
Eq.~(\ref{11}) indeed reproduces the Bose (Fermi) probabilities 
when $g=0$ $(g=1)$, but it can be negative for
certain values of $g\in]0,1[$. This should not be considered as 
problematic since exclusion statistics has to be understood for a
macroscopic ensemble of interacting particles  occupying a macroscopic
number of quantum states. In other words, one quantum state probability or
one quantum state grand partition function, if 
precisely defined formally,  have
 no real physical meaning, as well as any spectral quantity computed
from them. However, they become physically relevant  when 
integrated over the whole range of the energy spectrum, as we will see later.

Let us now follow the first quantized procedure used above in the Bose and
Fermi cases. The probabilities (\ref{prob}) should now become
\be\label{12} {x_1^{N_1}P_{N_1}\over \Xi_1} {x_2^{N_2}P_{N_2}\over \Xi_2}
{\cal{W}}^{N_1N_2}_{k_1k_2} \;.\ee
Again, one has $\sum^{\infty} 
{x_1^{N_1}P_{N_1}\over \Xi_1} {x_2^{N_2}P_{N_2}\over \Xi_2}
{\cal{W}}^{N_1N_2}_{k_1k_2}=1$.
The average number of transmitted carriers is
\bea\nonumber
\langle k \rangle & =&
\sum^{\infty}
{x_1^{N_1}P_{N_1}\over \Xi_1}
{x_2^{N_2}P_{N_2}\over \Xi_2}
{\cal{W}}^{N_1N_2}_{k_1k_2}(k_1-k_2)\\&=&\label{13} T(f_1-f_2)\;,
\eea
i.e. the finite temperature Landauer formula also holds for  
exclusion statistics. The 1d average current  then reads
\be\label{17} \langle I\rangle 
\equiv {q\over h}\int_0^{\infty}T(f_1-f_2)d\epsilon \;.\ee
By setting the voltage as $qV\equiv \mu_1-\mu_2$, the linear conductance 
follows as
\be\label{18} G\equiv \lim_{V\to 0}{\langle I\rangle \over V}={q^2\over h}T 
f_1 ({\epsilon=0}) \;, \ee
where $f_1(\epsilon=0)$ is the mean occupation number
at zero energy and chemical potential $\mu_1$ ($f_1(\epsilon=\infty)=0$).

Similarly, the spectral noise should be  defined from
\[
  \langle k^2 \rangle =
  \sum^{\infty}
  {x_1^{N_1}P_{N_1}\over \Xi_1}
  {x_2^{N_2}P_{N_2}\over \Xi_2}
  {\cal{W}}^{N_1N_2}_{k_1k_2}(k_1-k_2)^2 \;, 
\]
which yields
\bea\nonumber 
\langle k^2 \rangle - {\langle k \rangle}^2
=&T&(f_1+f_2) \\
\label{14} &+&T^2[f_1(C_1-1)+f_2(C_2-1)]\;, 
\eea
where  $C(x)\equiv  x{f'\over f}= (1-g f)[1+(1-g)f]$ is related
\cite{Raja}
to the fluctuation of the mean occupation number $\langle\delta f^2\rangle=fC$.
Again, one is confronted to the same difficulty as in the Bose and Fermi cases,
that is to say a term has to be added to (\ref{14})  to ensure that
both the fluctuation-dissipation theorem and quantum suppression at zero 
temperature hold. 
We believe that the fluctuation-dissipation theorem is valid for
exclusion statistics,  eventhough there is no definite proof that
it is indeed the case.
The fluctuation-dissipation theorem requires that when
$\mu_1=\mu_2$, the $T^2$ part of the noise should vanish.
In the Bose and Fermi cases,  the term added to (\ref{7}) is
proportional to $T(1-T)$ and symmetric in 1 and 2. Let us generalize this
construction to exclusion statistics by adding to the first
quantized spectral noise  (\ref{14}) a term proportional to $T(1-T)$,
symmetric in 1 and 2, and such that the
$T^2$ part disappears when $\mu_1=\mu_2$. These constraints uniquely
determine this term to be $ T(1-T)[f_1(C_2-1)+f_2(C_1-1)]$
which results in
\bea\nonumber \langle k^2 \rangle - {\langle k \rangle}^2
=&T&(f_1C_1+f_2C_2)\\ 
\label{16} &-&T(1-T)(f_1-f_2)(C_1-C_2)\;, \eea
  reproducing the Bose (Fermi) cases (\ref{9})  when $g=0$ ($g=1$).
The zero frequency spectral density of noise follows as
\bea\nonumber P=2{q^2\over h}\int_0^{\infty} &\bigg(&T(f_1C_1+f_2C_2)\\
\label{19} &-&T(1-T)
(f_1-f_2)(C_1-C_2)\bigg) d\epsilon \;.\eea
Does it satifies the fluctuation-dissipation theorem and quantum suppression?

(i) Fluctuation-dissipation theorem: (\ref{19})
yields when $\mu_1=\mu_2$ the thermal noise
\be\label{20}
P_{\rm thermal}=2{q^2\over h}2T\int_0^{\infty} f_1C_1d\epsilon=2{q^2\over h} 
{2T\over \beta}f_1(\epsilon=0)\; .\ee
Thus, from (\ref{18}),  the relation $P_{\rm thermal}=4G/\beta$  holds for 
exclusion statistics, which  manifests the fluctuation-dissipation theorem.

(ii) Quantum suppression of the shot noise: 
at zero temperature on the other hand, 
only the $T(1-T) $ term contributes to the noise
(\ref{19}). Since at zero temperature
\be\label{21} \int_0^{\infty}(f_1-f_2)(C_1-C_2)d\epsilon=
-{1\over g}(\mu_1-\mu_2)\;, \ee
one obtains the shot noise
\be\label{22} P_{\rm shot}=2{q^2\over h}{1\over g}T(1-T)(\mu_1-\mu_2)\ee
which is indeed proportional to $T(1-T)$. Note also that when 
$g\to 0$, (\ref{21},\ref{22}) become meaningless, a mere reflection of the
absence of a well defined Bose spectral noise at small temperature.

{\it Crossover from thermal to shot noise.---}
Let us now assume   that 
$e^{\beta\mu_1}\simeq e^{\beta\mu_2} \gg 1$, 
even though ${\beta(\mu_1-\mu_2)}$ can
be  small (thermal noise) or large (shot noise)  \cite{Oriol}. 
The question we now ask is what is the crossover
regime which interpolates between these two situations?
 
When  $e^{\beta\mu_{1}}, \;e^{\beta\mu_{2}} \gg1$, one has that not only
$f(\epsilon=0)=1/g$ but also 
$\int_0^{\infty} fd\epsilon={\mu/g}$. Then in the shot noise limit 
the  current (\ref{17}) and the conductance  (\ref{18}) become
\be\label{23} \langle I\rangle ={q\over h}T{\mu_1-\mu_2\over g}\;, \quad
G={q^2\over h}{1\over g}T \;.\ee
The conductance in (\ref{23}) is a generalization to exclusion statistics 
of the usual Landauer
conductance. Comparing (\ref{22}) and (\ref{23}), one has
\be\label{24} P_{\rm shot}= 2q \langle I\rangle (1-T)  \;. \ee
We stress that the shot noise ({\ref{24}})
has a universal form independent of the statistics of charge carriers.
When $T\to 0$, the Schottky limit
$ P_{\rm shot}= 2q \langle I\rangle$ is obtained. It was
used to measure the fractional charge $q=e/3$ of the carriers in
$\nu=1/3$ FQHE devices \cite{Glattli}. Remarkably, 
the quantum suppression factor $1-T$ was also confirmed
experimentally when $T$ is not small \cite{Glattli}  for charge  carriers 
which are supposed to obey fractional or exclusion statistics.
Note that the average current $\langle I \rangle$ transmitted through the 
scattering region should be identified with the 
tunneling (or backscattering) current in the experiments \cite{Glattli,Nature}.

If we  assume that the charge carriers are 1d
quasiholes obeying exclusion statistics and
carrying a charge $q=ge$
then the  conductance  in (\ref{23}) becomes
$G={e^2\over h} g T  $,
the FQHE conductance for the tunneling  current.

Keeping
$e^{\beta\mu_{1}},\;e^{\beta\mu_{2}} \gg 1$, the energy integration of  
the spectral noise (\ref{19}) can be simplified to
\be\label{26} P={2q^2\over h}{1\over g}{2\over \beta}\left[T +T(1-T)
\delta P\right]\;, \ee
where the dimensionless excess noise 
\be\label{27} \delta P
=g\int_0^{\infty}{[f(a^2x)+f(a^{-2}x)]C(x)\over 2x}{dx}-1 \;,\ee
which only depends on  the parameter $a\equiv \exp(\beta qV/2)$,
represents the noise in units of $4G(1-T){/ \beta}$.
Note that $\delta P\to 0$ when $\beta qV/2\to 0$ (thermal noise), and
$\delta P\to \beta qV/2$ when $\beta qV/2\to \infty$ (shot noise), 
as it should.

In the Fermi case $g=1$, with $q=e$, (\ref{26},\ref{27}) yield the 
standard result
\be\label{28}  P={2e^2\over h}{2\over \beta}\left[T +T(1-T)\bigg(
{\beta eV\over 2}\coth{\beta eV\over 2}-1\bigg)\right] \;.\ee
In the case $g=1/2$, where $f(x)=x/\sqrt{1+x^2/4}$, one  obtains
if $a<1$
\[ 
\delta P={\arctan
\sqrt{{a^{-4}}-1}
\over 2\sqrt{{a^{-4}}-1}}- {1\over 4\sqrt{1-{ a^4}}}
\ln \frac{1- \sqrt{1-a^4}}{1+\sqrt{1-a^4}}
-1 \;,    
\]
and if $a>1$, $a\to{ 1/a}$ in the above expression. 
In the case of interest $g=1/3$, i.e. fractional statistics quasihole in
FQHE devices at filling factor $\nu=1/3$, one has
\[
f(x)={3\over 2}(w+{y\over w}),\;\;
 w^3\equiv {y}\bigg(1+\sqrt{1-y}\bigg),
\;\; y\equiv {{4\over 27}x^3\over 1+{4\over 27}x^3}. 
\] 
 The thermal and shot noise rewrite respectively as
\be\label{30} P_{\rm thermal}={4\over 3}{e^2\over h}T{1\over\beta}\;, 
\quad  P_{\rm shot}={2\over 3}{e^2\over h}T(1-T){eV\over 3}\;. \ee

The crossover has been studied experimentally \cite{Glattli,Nature} at
$\nu=1/3$. In \cite{Nature},  by replacing
in the fermion result (\ref{28}), the quantum of conductance
$e^2/h\to e^2/(3h)$ and
$\beta e V\to \beta eV/3$, a reasonably good agreement with
the  experimental data could be obtained. These two changes
happen to be fine tuned such that the thermal and
shot noises coincide with (\ref{30}). We propose here
to go beyond this heuristic approach and compare the experimental data
with the exact crossover formula (\ref{26},\ref{27})
when $g=1/3$ and $q=e/3$.
The dimensionless excess noise (\ref{27}),
being a function of a single variable $\beta q V /2$,
can easily be analyzed numerically.
In principle, the crossover from the shot to thermal noise
can be  achieved in two ways:
either by varying the voltage $V$ at  fixed temperature (the standard
experimental setting
to look at the crossover  as a function of the backscattering current
 \cite{Nature}), or vice versa.
In the former case, the dimensionless excess noise (\ref{27}) is well suited.
Note  in particular the asymptotic behavior when $\beta q V/2
\to \infty$ (shot noise)
$ \delta P- {\beta qV\over 2} =c(g)-1 + \ldots  $
where $c(g)$ depends only on $g$ and the dots represent terms vanishing
as $\beta q V/2\to \infty$. One has
$c(1)=0$, $c(1/2)=\ln 2$ and we found numerically that
$ c(1/3) \approx  0.5493 $ from which one can infer that
$c(1/3)=(\ln 3)/2$. In the shot noise limit, this subleading constant term
contains information about the statistics of the carriers.

When the crossover is achieved by varying the temperature at a fixed voltage,
the dimensionless excess noise
$ {\delta P'} =  (2/\beta q V) \delta P  $,
which counts  the noise in units of $2G (1-T) qV$, is more appropriate. It
is displayed in Fig. 1 for $g=1/3$. 
%along with the modified fermion expression.
\begin{figure}
\centerline{\epsfig{file=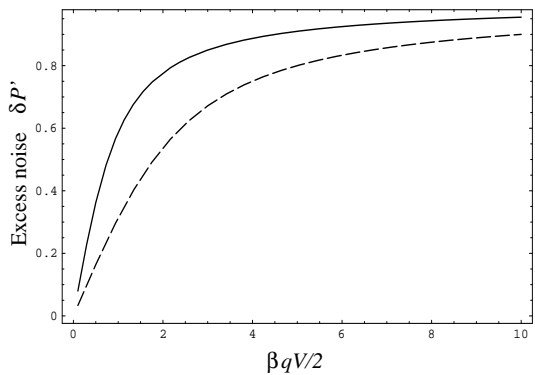,angle=0,width=7cm}}
\caption{The dimensionless crossover function $\delta P'$ for $g=1/3$ and
$q=e/3$ representing the
excess noise in units of $2{(e^2/ 3h)}T(1-T) eV/3$
as a function of $\beta q V/2$ (solid line).  
The same for the fermionic excess noise with
 $e^2/h\to e^2/(3h)$ and  
$\beta e V\to \beta eV/3$ as in [2] %\cite{Nature} 
(dashed line).} 
\label{fig:crossover}
\end{figure}
\noindent 
The difference between the two curves is maximal for $\beta q V/2\simeq 1$
and vanishes when $\beta q V/2\to 0,\infty$.

As to tunneling experiments in FQHE devices, the thermal noise limit 
is eventually fixed by the fluctuation dissipation theorem
$P_{\rm thermal}=4G/\beta$, where $G=\nu e^2/h$ is the FQHE conductance. 
On the other hand,
the quantum shot noise  has the  universal form (\ref{24}) independent of the
statistics of the charge carriers. The statistics of charge carriers
can be experimentally determined by analyzing the crossover between
these two limits. This should in particular 
allow for an unambiguous experimental discrimination
between  fractional charge carriers with Fermi   statistics and
fractional charge carriers with exclusion statistics \cite{final}.

We would like to thank M. B\"uttiker and 
C. Glattli for discussions.

\end{multicols}

\begin{references}
\bibitem{Oriol} The experiments \cite{Glattli,Nature} are  performed  at 
low temperature, of the order of 10-150 mK.
\end{references}
\end{document}